\documentclass[conference]{IEEEtran}
\IEEEoverridecommandlockouts
\usepackage{cite}
\usepackage{amsmath,amssymb,amsfonts}
\usepackage{algorithmic}
\usepackage{graphicx}
\usepackage{textcomp}
\usepackage{xcolor}
\usepackage{algorithm}
\usepackage{diagbox} 
\usepackage{booktabs}
\usepackage{tabularx}
\usepackage{soul}

\def\BibTeX{{\rm B\kern-.05em{\sc i\kern-.025em b}\kern-.08em
    T\kern-.1667em\lower.7ex\hbox{E}\kern-.125emX}}
\begin{document}

\title{Efficient Layered New Bit-Flipping QC-MDPC Decoder for BIKE Post-Quantum Cryptography\\
}

\author{
\IEEEauthorblockN{Jiaxuan Cai and Xinmiao Zhang}
\IEEEauthorblockA{Dept. of Electrical \& Computer Engineering, The Ohio State University, Columbus, OH 43210 U.S.A. \\
\{cai.1072, zhang.8952\}@osu.edu}
}

\maketitle

\begin{abstract}
The medium-density parity-check (MDPC) code-based Bit Flipping Key Encapsulation (BIKE) mechanism remains a candidate of post-quantum cryptography standardization. The latest version utilizes a new bit-flipping (BF) decoding algorithm, which decides the BF threshold by an affine function with high-precision coefficients. Previous BF decoder implementations can be extended to the new algorithm. However, they suffer from large memories that dominate the overall complexity. This paper proposes a column-layered decoder for the new BIKE BF decoding algorithm to substantially reduce the memory requirement, and optimizes the affine BF threshold function coefficients to reduce the code length needed for the same security level. For the first time, our work also investigates the impact of finite precision representation of the threshold coefficients on the decoding performance. For an example MDPC code considered for the standard, the proposed layered BF decoder achieves 20\% complexity reduction compared to the best prior effort with a very small latency overhead.

\end{abstract}

\section{Introduction}

With the advancement of quantum computers, the National Institute of Standards and Technology (NIST) launched the post-quantum cryptography standardization to develop cryptographic schemes that can withstand quantum attacks \cite{NIST}. The Bit Flipping Key Encapsulation (BIKE) mechanism based on quasi-cyclic medium-density parity-check (MDPC) codes \cite{BIKE} is a finalist in the latest submission round.

The bit-flipping (BF) decoding algorithm \cite{BF} and its variations \cite{Liva, Baldi, GuneysuDATE, Guneysu, Hu, XieSparse} have been extensively studied for decoding MDPC codes, which is a critical step in the decapsulation phase of BIKE. In BF algorithms, if the number of nonzero syndromes participating a column of the parity check matrix exceeds a threshold, the corresponding bit is flipped. Implementations in \cite{Liva, Baldi, GuneysuDATE, Guneysu, Hu, XieSparse} divide each column of the MDPC parity-check matrix into $L$-bit segments and process the segments sequentially. In \cite{XieSparse}, the sparsity of parity-check matrices is taken into account and only the non-zero segments are processed to reduce the latency. Compared to BF algorithms, the Min-sum algorithm \cite{Minsum} can substantially improve the decoding failure rate (DFR) at the cost of higher complexity. Min-sum decoding algorithm optimizations for MDPC codes and parallel decoder implementation architectures have been proposed in \cite{SMPMS, RLMS, LLMS}.

The BIKE scheme previously utilizes the Black-Gray-Flip (BGF) decoding algorithm \cite{BGF}, which introduces two auxiliary steps to lower the DFR with a small number of iterations. However, its security is compromised by weak key attacks \cite{weakkey}. To address this issue, the new BIKE BF decoding algorithm \cite{newAlgo} was proposed. By adjusting the bit-flipping threshold over the decoding iterations, the new BF algorithm also achieves better DFR and accordingly requires shorter codes to achieve the same level of security.

While the BGF decoder implementation \cite{foldingBIKE} can be adapted for the new BF algorithm, it requires memories to store the syndromes for both the previous and current decoding iterations and duplicates these memories for each iteration to enable parallel processing. These memories contribute to a significant portion of the decoder complexity. Additionally, the thresholds in BIKE BF decoders are computed using affine functions with high-precision coefficients. However, finite precision representation is essential in hardware implementation in order to reduce the datapath and hardware complexity.

This paper proposes a parallel column-layered decoder for the new BIKE BF decoding algorithm to substantially reduce the memory requirement. Only the syndromes for one decoding iteration are stored and the updated syndromes are written back in place. Besides, the syndromes can be extracted from even and odd memory banks without duplicating the memory. Various coefficients for BF threshold computation are investigated to reduce the DFR and accordingly the codeword length for achieving the same security level. For the first time, the impact of finite precision representation of the threshold coefficients is investigated, revealing that it significantly affects the DFR. However, a moderate number of fractional bits can be kept to make the performance loss negligible and hardware implementation data path short. For an example MDPC code used to achieve 128-bit security, the proposed layered BF decoder reduces the hardware complexity by 20\% with a very small latency overhead compared to the best prior effort.

\section{Background}

\begin{figure}[t] 
    \begin{center}
    \includegraphics[width=.4\textwidth]{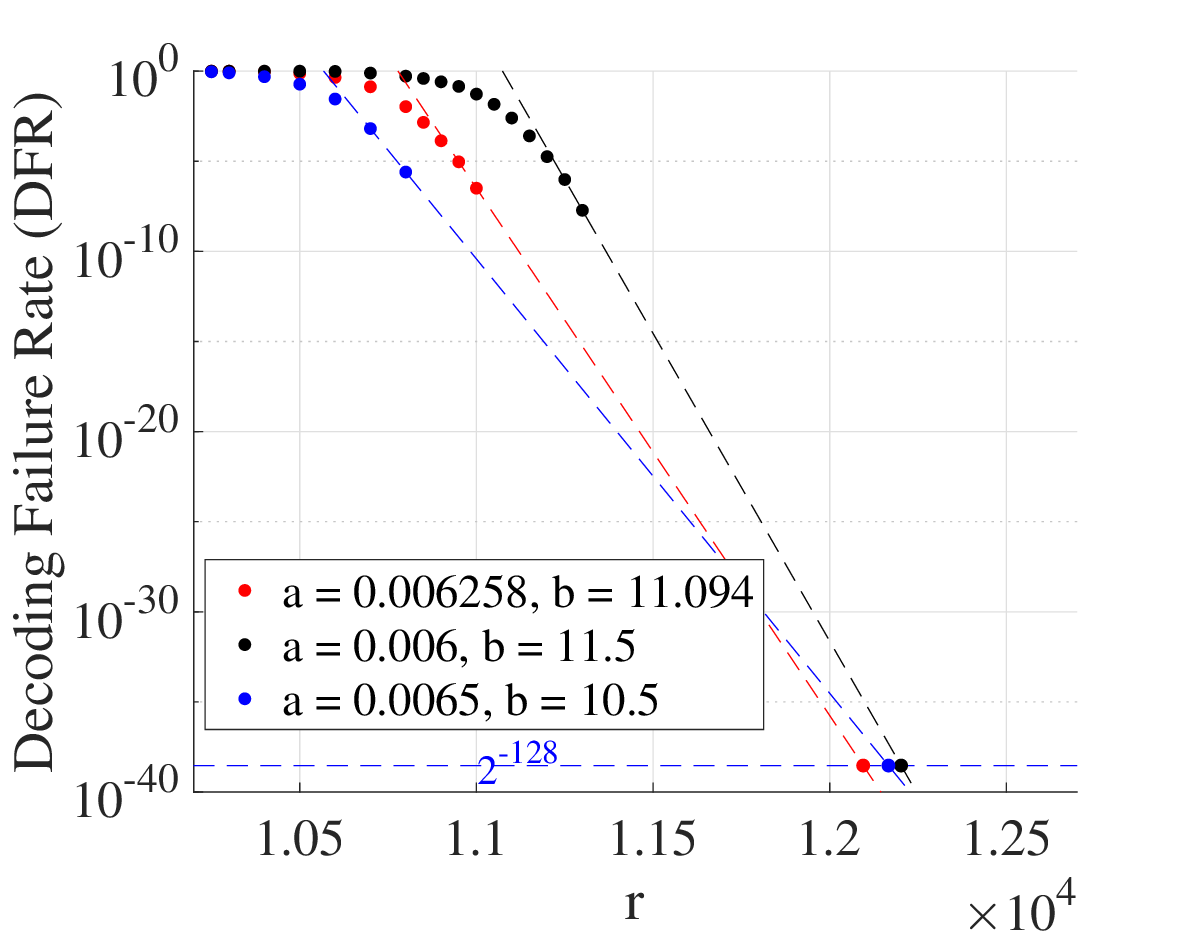}
    \vspace {-0.5em}\caption{DFRs of the new BIKE BF decoder with different threshold coefficients and the extrapolation of  $r$ required by $\lambda=128$ for MDPC codes with $(w, t)=(142,134)$ and $I_{\text {max}}=7$.}\label{extrapolation}
    \end{center}
    \vspace {-1em}
\end{figure}

The MDPC code used in the BIKE mechanism is characterized by its parity-check matrix $\mathbf{H} = [\mathbf{H}_0 | \mathbf{H}_1]$, consisting of two circulant submatrices. During key generation step of BIKE, two vectors, $h_0$ and $h_1$, each with $r$ bits and Hamming weight $d = w/2$, are randomly generated and are used as the first rows of $\mathbf{H}_0$ and $\mathbf{H}_1$, respectively. The other rows of $\mathbf{H}_0$ and $\mathbf{H}_1$ are generated by cyclically shifting their first rows. If $\mathbf{H}_0$ is non-invertible, $h_0$ must be randomly regenerated. The matrix $\mathbf{H}$ serves as the private key, while $\mathbf{H}_1\mathbf{H}_0^{-1} $ is used as the public key. In the encapsulation phase, two vectors $e_0$ and $e_1$ such that $|e_0| + |e_1| = t$ are randomly generated to compute $c = e_0 + e_1 (\mathbf{H}_1 \mathbf{H}_0^{-1})^T$. During the decapsulation, the syndrome vector is computed as $s = c\mathbf{H}_0^T$, then MDPC decoding is performed on $s$ to recover $e=[e_0|e_1]$ \cite{BIKE}. For MDPC codes used in BIKE, the values of $w$ are 142, 206, and 274, and those of $t$ are 134, 199, and 264 for security level $\lambda=128, 192$, and $256$, respectively.

\begin{algorithm}
\caption{New BIKE BF Decoding Algorithm \cite{newAlgo}}
\label{newAlgo}
\begin{algorithmic}[1]
    \STATE {{\bf Input:} $s, \mathbf{H}$}; {\bf Output: $e$} 
\STATE{\bf Function for optimal BF threshold:} {$f(x)=ax+b$} 
    \STATE{\bf Parameters:} {$a, b,\delta$, which are dependent on $\lambda$}
\STATE{{\bf Initialization:} $e \gets 0$}  
\vspace{0.5em}
  \STATE $T' \gets f(|s|)$; $M \gets (d+1)/2$
    \STATE {\bf for} {$i=1$ to $I_{max}$} {\bf do}
    \STATE $\quad$ $T \gets \texttt{THRESHOLD}(i,{|s|})$; $\tilde s\gets s$
    \STATE $\quad$ {\bf for} {$j=0$ to $2r-1$} {\bf do}
    \STATE $\quad$ $\quad$ $\sigma_j \gets \texttt{count}(\mathbf{H},{s},j)$
    \STATE $\quad$ $\quad$ {\bf if} $\sigma_j \geq T$ {\bf then}
    \STATE $\quad$ $\quad$ $\quad$ $e_j \gets e_j \oplus 1$; $\Tilde{s} \gets \Tilde{s} \oplus \texttt{col}(\mathbf{H},j)$
\STATE $\quad$ $s\gets \tilde s$
    \STATE {\bf if} $ s=0$ {\bf then} {\bf return} $e$; {\bf else} {declare decoding failure}
\vspace{0.5em}
    \STATE {\bf function} $\texttt{THRESHOLD}(i,{|s|})$
    \STATE $\quad$ {\bf if} $i \{=1, =2, =3, \geq4\}$
 \STATE $\quad$ {\bf then} $T\! \gets\! \{T'\!+\!\delta, (2T'\!+\!M)/3 \!+\! \delta, (T'\!+\!2M)/3\! +\! \delta, M + \delta\}$
    \STATE $\quad$ {\bf return} $\texttt{max}(f(|{s}|),T)$
\end{algorithmic}
\end{algorithm}

The pseudo code for the new BIKE BF decoding algorithm \cite{newAlgo} is provided in Algorithm \ref{newAlgo}. This algorithm takes $\mathbf{H}$ and the syndrome vector $s$ as input, and outputs the vector $e$. The coefficients $a$ and $b$ for the affine threshold function, $f(x)$, are derived from least squares fitting\cite{Vasseur} and the optimal values change with $\lambda$. The optimal value of positive integer, $\delta$, is also decided by the security level, $\lambda$ \cite{newAlgo}. The $\texttt{THRESHOLD}$ function decides the threshold for bit flipping for each decoding iteration based on the weight of the syndrome vector in the beginning of the iteration. The function $\texttt{count} (\mathbf H, s, j)$ counts the number of nonzero syndromes participating in column $j$ of $\mathbf H$. If this count exceeds the threshold value $T$, the corresponding bit is flipped and the syndrome vector is updated by XORing column $j$ of $\mathbf H$, denoted by $\texttt{col}(\mathbf H,j)$. To mitigate timing attacks, the decoding is always run through $I_{max}$ iterations. At the end, the vector $e$ is returned if the syndrome vector becomes zero.

To achieve $\lambda$ bits of security, the indistinguishability under chosen ciphertext attack (IND-CCA) requires the DFR to be under $2^{-\lambda}$ \cite{BIKE}. The DFR decreases as $r$ increases. Fig. \ref{extrapolation} depicts the DFR vs $r$ curves of the new BIKE BF decoder with various $a$, $b$ and $\delta=3$ \cite{newAlgo} for an MDPC code targeting at $\lambda=128$ bits of classical security. Simulations at low DFR can not be completed within a reasonable amount of time. Instead, the curves at lower DFRs are extrapolated by curve fitting on the two lowest DFR values derived from simulations. To reduce the key size and implementation complexity, the setting with the smallest $r$ at DFR=$2^{-128}$ should be chosen. It was found in \cite{newAlgo} that $a=0.006258$ and $b=11.094$ lead to minimal $r$ for the new BIKE BF decoding of the $(w,t)=(142, 134)$ MDPC code. Slightly changing $a$ and $b$ leads to significant differences in the DFRs and resulting $r$.

\section{Layered Finite-Precision BIKE BF Decoder}

This section presents a column-layered BIKE BF decoder and explores various coefficients for BF threshold decision to improve the DFR. Then the impacts of finite precision representation of the threshold coefficients are investigated.

\subsection{Layered BIKE BF Decoder}

The decoder design in \cite{foldingBIKE} for the BIKE BGF algorithm can be modified to implement Algorithm \ref{newAlgo}. Over 90\% of the decoder silicon area is contributed by memories. As shown in Line 9 of Algorithm \ref{newAlgo}, the syndromes in the beginning of a decoding iteration are always used to count the number of nonzero participating syndromes for each column of $\mathbf H$.  However, the syndromes need to be updated after each bit flipping in Line 11. As a result, two memories are required to store $s$ and $\tilde s$, and they account for around a half of the overall decoder memories. To address this issue, this paper proposes to adopt column-layered scheduling for the BIKE BF decoding. The updated syndromes from Line 11 of Algorithm \ref{newAlgo} are used right away in  counting the nonzero participating syndromes for the next column in Line 9. In this case, the syndromes can be updated in place and only one memory is needed for the syndromes.

\begin{figure}[t] 
    \begin{center}
    \includegraphics[width=.4\textwidth]{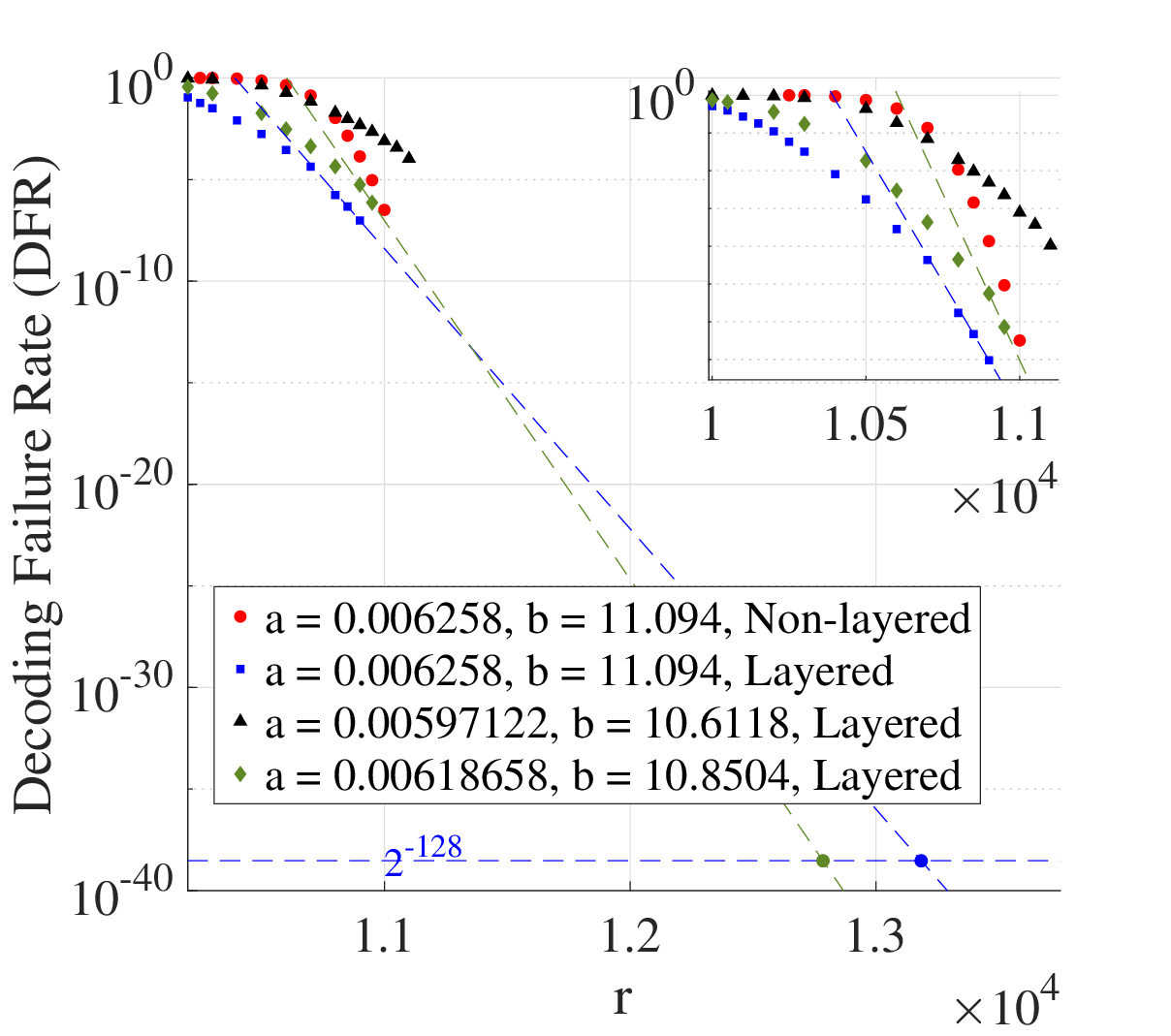}
    \vspace {-1em}\caption{DFRs of optimal non-layered decoder and layered decoders with different $a$ and $b$ values for MDPC codes with $(\lambda, w, t)=(128, 142,134)$ and $I_{\text {max}}=7$. }\label{DFR1}
    \end{center}
    \vspace {-1.5em}
\end{figure}

\begin{table}[t]
   \renewcommand\arraystretch{1.15}
   \caption{$r'$ and Corresponding $a$ and $b$ Values for Layered BIKE BF Decoder for MDPC Codes With $(\lambda,w,t)=(128, 142, 134)$}\label{layer_optimize} \vspace{-1em}
    \begin{center}
    \begin{tabular}{c|c|c||c|c|c}
    \hline
    $r'$ &$a$ & $b$ & $r'$ & $a$ & $b$ \\\hline\hline
    11000 & 0.00622942 &11.4157 &11300 & 0.00590374 & 10.2409\\\hline
    11100 & 0.00618658 &10.8504 &11400 & 0.00586073 & 10.1113\\\hline
    11200 & 0.00597122 &10.6118 &11500 & 0.00577619 & 9.9775\\\hline
    \end{tabular}
    \end{center}
    \vspace {-2.5em}
\end{table}

The optimal values of the BF threshold coefficients, $a$ and $b$, are highly dependent on the decoding algorithm. The threshold coefficients optimized for the non-layered decoding in Algorithm \ref{newAlgo} may not be the best for our proposed layered decoding. Inspired by the procedure for deciding the optimal $a$ and $b$ in \cite{BIKE, Vasseur}, multiple values of $r'$ are first chosen as shown in Table \ref{layer_optimize} for MDPC codes with $(\lambda, w, t) = 128,142,134$. These $r'$ are selected based on the $r'$ values given for the BGF decoder and the fact that the $r'$ for Algorithm \ref{newAlgo} should be larger than that for the BGF algorithm \cite{Vasseur}. For each $r'$, a large number of random $r'\times 2r'$ $\mathbf H$ matrices and $2r'$-bit $e$ vectors are generated and decoding is carried out. In each decoding instance, every integer ranging from 30 to 60 is tried as the BF threshold and the threshold leading to the largest reduction on the syndrome weight after the first decoding iteration is chosen. Plotting the chosen thresholds and initial syndrome weights in a two-dimensional space, a pair of $a$ and $b$ values can be derived by least square fitting as listed in Table \ref{layer_optimize}. 

For each $a$ and $b$ pair, the DFR curve can be plotted by extrapolating simulation results. For conciseness, the results for some $a$ and $b$ pairs are shown in Fig. \ref{DFR1}. The pair leading to the smallest $r$ at DFR=$2^{-128}$, $a=0.00618658$ and $b=10.8504$, is selected as the optimized BF threshold coefficients. It can be observed that this pair leads to much smaller $r$ than the pair optimal for the non-layered decoding. It should be noted that better $a$ and $b$ pairs may be found by starting with different $r'$ or using alternative procedures.


\subsection{Decoder with Finite Precision Representation}

\begin{figure}[t] 
    \begin{center}
    \includegraphics[width=.4\textwidth]{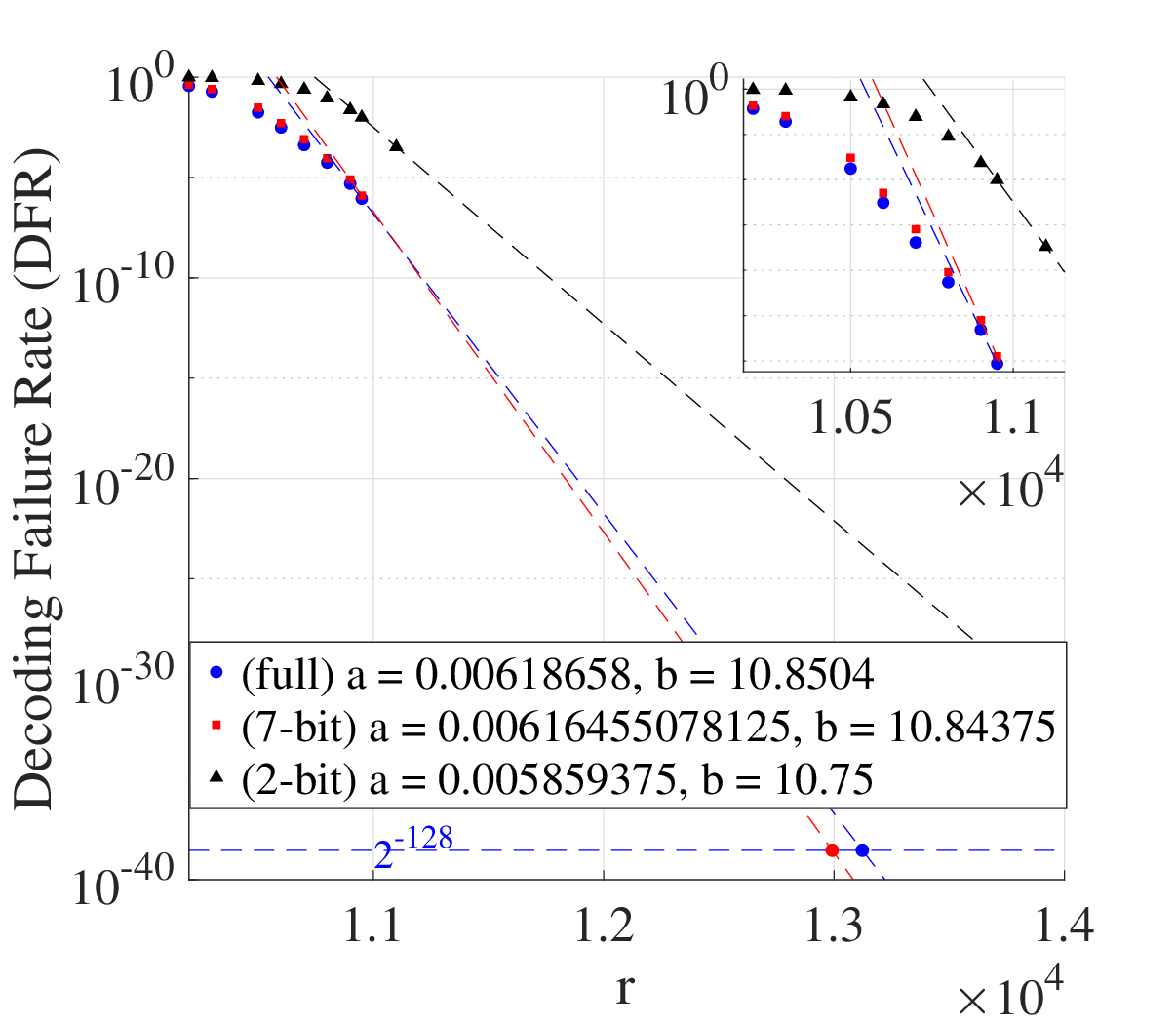}
    \vspace {-1em}\caption{DFRs of parallel layered BF decoder with different precision in $a$ and $b$ for MDPC codes with $(\lambda, w, t)=(128, 142,134)$, $L$=32, and $I_{\text {max}}=7$.}\label{DFR2}
    \end{center}
    \vspace {-1.5em}
\end{figure}

\begin{table}[t]
   \renewcommand\arraystretch{1.15}
   \caption{Binary Representations and Values of the Coefficients With Different Precision Levels for Layered Decoder}\label{fin_pre} \vspace{-1em}
    \begin{center}
    \begin{tabular}{@{}c@{}|c|@{}c@{}|@{ }c@{}}
    \hline
     precision & coeff. & binary & decimal \\\hline\hline
     full & a &0.00000001100101010111$\cdots$ &0.00618658 \\\cline{2-4}
     & b & 1010.11011001101100111101$\cdots$&10.8504 \\\hline
     7 bits & a &0.00000001100101 & 0.00616455078125\\\cline{2-4}
     & b &1010.1101100&10.84375 \\\hline
     2 bits & a &0.000000011 & 0.005859375\\\cline{2-4}
     & b &1010.11 &10.75 \\\hline
    \end{tabular}
    \end{center}
    \vspace {-2.5em}

\end{table}

The optimal $a$ and $b$ need many bits to represent. The multiplication and addition with high-precision numbers lead to long data path and large area. To reduce the hardware complexity, the lower bits of the coefficients may be discarded. On the other hand, the truncation may affect the DFR. 

Converting the optimal $a$ and $b$ found from the aforementioned procedure leads to 60 and 51 bits, respectively, in the binary representations of the fractional parts as listed in Table \ref{fin_pre}. Our work explores two finite-precision designs. One keeps the 2 most significant nonzero bits (MSNBs) of $a$ and 2 most significant bits (MSBs) of $b$ in the fractional parts, and the other keeps the 7 MSNBs of $a$ and 7 MSBs of $b$. The DFRs corresponding to different precision are plotted in Fig. \ref{DFR2}. When 7 bits are kept in the fractional parts, the DFR and accordingly extrapolated $r$ only increase slightly from the case with full precision. However, keeping only 2 bits in the fractional parts lead to significant degradation. 

\section{Hardware Architectures and Comparisons}

\begin{figure}[t] 
    \begin{center}
    \includegraphics[width=.485\textwidth]{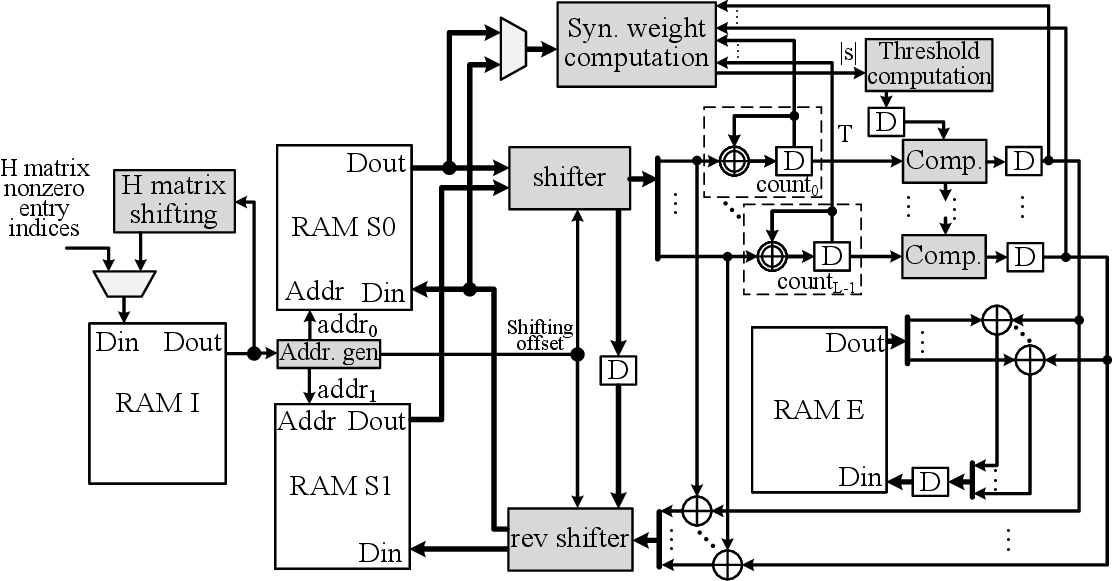}
    \vspace {-1.5em}\caption{Top-level block diagram of the proposed layered new BF decoder.}\label{arch}
    \end{center}
    \vspace {-1.5em}
\end{figure}

\begin{table}[t]
   \renewcommand\arraystretch{1.15}
   \caption{Sizes of the RAMs Utilized in the Proposed $L$-Parallel BIKE BF Decoder and Best Prior Non-Layered Decoder for $(r,w)$ MDPC Codes}\label{rams} \vspace{-1em}
    \begin{center}
    \begin{tabular}{c|c|c}
    \hline
     &proposed layered decoder & best prior non-layered decoder \\\hline\hline
    RAM E &  $\lceil 2r/L \rceil \times L$  & $\lceil 2r/L \rceil \times L$ \\\hline
    RAM S & $2\times \lceil r/2 L\rceil \times L$ &$4\times \lceil r/2 L\rceil \times L$\\\hline
    RAM I &$2 \lceil log_2(r) \rceil \times w/2$ & $2 \lceil log_2(r)\rceil \times w/2$\\\hline
    \end{tabular}
    \end{center}
    \vspace {-1.5em}
\end{table}

This section first proposes an efficient parallel layered BIKE BF decoder architecture. Then the hardware complexity is analyzed and compared with that of the best prior design.

Fig. \ref{arch} presents the top-level block diagram of the proposed $L$-parallel layered BIKE BF decoder, which processes $L$ columns of $\mathbf{H}$ simultaneously. To minimize the memory requirement, only the row indices of the nonzero entries in one column for each submatrix of $\mathbf H$ are stored in RAM I. Each index is used to process a diagonal of $L$ entries in a block of $L$ columns in each clock cycle. The row indices for processing the subsequent blocks of $L$ columns are derived by adding $L$ mod $r$ in the ``$\mathbf{H}$ matrix shifting'' module. RAM S stores the most updated syndromes of $L$ consecutive rows in each address. It consists of two banks: S0 and S1, and they store the syndromes for even and odd block rows, respectively. Although the $L$ diagonal entries to process in each clock cycle could start from any row and they do not necessarily align with the $L$ rows stored in a single RAM S location, they can always be extracted from two adjacent addresses of RAM S, one in RAM S0 and the other in RAM S1, by the shifter architecture in \cite{RLMS}. The RAM E stores $L$ bits of the updated vector $e$ in each address. The addresses of both RAM E and RAM I are generated by counters. Table \ref{rams} summarizes the sizes of the RAMs used in the proposed $L$-parallel layered decoder. 
For comparison, the memory requirement of the non-layered BIKE BF decoder is also listed in this Table. Unlike the design in \cite{foldingBIKE}, which uses two copies of RAM S to enable the access of arbitrary $L$ consecutive syndromes at a time, it is assumed that the even-and-odd bank design of RAM S is adopted to implement the non-layered design. However, it needs to store both $s$ and $\tilde s$ and hence has twice the size in RAM S compared to our proposed design.

Our $L$-parallel decoder employs $L$ counters and $L$ comparators. In each clock cycle, the $L$ syndromes corresponding to a $L$-bit diagonal of $\mathbf H$ extracted by the shifter are sent to the counters. After all the diagonals in a block of $L$ columns are processed, $\sigma_j$ are derived and they are compared with the threshold $T$ computed by the ``Threshold computation'' block. The comparator outputs indicate whether the bits should be flipped. Going through the same block of $L$ columns of $\mathbf H$ again, the bits in $e$ are flipped and the syndromes are updated as listed in Line 11 of Algorithm \ref{newAlgo}. The updated $e$ and syndromes are written back to the memories in place.   

At the beginning of decoding, the syndrome weight, $|s|$, for threshold computation is calculated by the ``Syn. weight computation'' block. It has an adder-register feedback loop that accumulates in each clock cycle the number of nonzero bits in an $L$-bit syndrome segment read from RAM S. To reduce the complexity, each segment can be divided into smaller groups. The numbers of nonzero bits in the groups are derived by combinational logic and then are added up. In later decoding iterations, if bit $j$ is flipped, $|s|$ is updated to $|s|+d-2\sigma_j$. 
\begin{table}[t]
   \renewcommand\arraystretch{1.15}
   \caption{Complexity and Latency Comparisons for the $32-$Parallel Proposed BIKE BF Decoder and Best Prior Non-Layered Decoder for MDPC Codes With $(w,t)=(142, 134)$}
\label{complexity} \vspace{-1em}
    \begin{center}
    \begin{tabular}{c|c|c}
    \hline
     &proposed layered  & best prior non-   \\
     &  decoder &   layered decoder \\\hline\hline
    $r$ value &12992&12095\\\hline
    RAM E & 25984 &24192 \\\hline
    RAM S & 12992 &24192\\\hline
    RAM I & 1988 &1988\\\hline
    Total memory (bits)& 40964 &50372\\
    (normalized) &(0.81)&(1)\\\hline
    Logic (\# of XORs)& 3780 &5134\\\hline
    Total area (\# of XORs)& 34503 &42913\\
    (normalized) &(0.80)&(1)\\\hline\hline
    Latency (\# of clk cycles) & 115304& 107352 \\
    (normalized) &(1.074)&(1)\\\hline
    
    \end{tabular}
    \end{center}
    \vspace {-2.5em}
\end{table}

As an example, the complexity of the proposed layered decoder with $L=32$ is listed in Table \ref{complexity}. From Fig. \ref{DFR2}, our proposed design needs to use $r=12992$ to achieve $\lambda=128$-bit security. The sizes of the memories in the decoder can be computed from Table \ref{rams}. 7-bit $a$ and $b$ representation is adopted in our design, and the complexities of the constant multiplier and adder used for the $f(x)$ function in the ``Threshold computation'' are estimated accordingly.  The shifter and reverse shifter are implemented by 6 rows of multiplexers \cite{RLMS}. The complexity of all logic units in our design in terms of the number of XOR gates is listed in Table \ref{complexity}. Assuming that the area required to store a bit in memory is equivalent to 0.75 XOR gates \cite{XieSparse}, the overall decoder complexity in terms of the equivalent number of XOR gates is estimated as in Table \ref{complexity}. For comparison, the complexity of a non-layered decoder using the best available design for each component is also listed in this Table. It needs a slightly smaller $r$ to achieve 128-bit security and hence requires RAM S of almost twice the size. As a result, the proposed design achieves 20\% area reduction. For different parallelisms, since the memory is dominating the overall complexity, our proposed design would still achieve substantial complexity reduction.

The prior non-layered decoder uses full precision representation of $a$ and $b$, which increases the logic complexity and datapath. However, the multiplication of $a$ and addition by $b$ are not in a feedback loop and hence can be pipelined. After pipelining, the critical paths in both the proposed and prior non-layered decoders lie in the comparators, consisting of 8 levels of logic gates. As a result, the two decoders can achieve the same clock frequency. Our layered decoder has 7\% increase in latency because of the larger $r$ value needed. 

\section{Conclusions}
This paper proposes a column-layered decoder for the new BIKE BF decoding algorithm. The proposed design substantially reduces the memory requirement and accordingly the overall complexity of the decoder. Various threshold computation function coefficients are explored to minimize the dimension of the MDPC code that is required to achieve the same level of security. Besides, for the first time, this paper investigates the impacts of the finite-precision representation of the threshold coefficients on the decoding performance. For an example decoder, our proposed design achieves 20\% complexity reduction with only 7\% latency overhead. Future work will further reduce the decoding latency.

\vspace{12pt}

\end{document}